\documentclass{aa}

\begin{document}    
    
    
\title{Cosmic ray acceleration at supergalactic accretion shocks: \\
a new upper energy limit due to a finite shock extension}
 
\author{M. Ostrowski \and G. Siemieniec-Ozi\c{e}b{\l}o}
 
\institute{Obserwatorium Astronomiczne, Uniwersytet Jagiello\'nski,
                   ul. Orla 171, 30-244 Krak\'ow, Poland}
 
\offprints{G. Siemieniec-Ozi\c{e}b{\l}o, e-mail: grazyna\@@oa.uj.edu.pl}
    
\date{Received 17 October 2001 / Accepted 19 February 2002}

\titlerunning{The cosmic ray upper energy limit at supergalactic
accretion shocks}

\authorrunning{M. Ostrowski \& G.Siemieniec-Ozi\c{e}b{\l}o}
    
\abstract{
Accretion flows onto supergalactic-scale structures are accompanied with
large spatial scale shock waves. These shocks were postulated as
possible sources of ultra-high energy cosmic rays. The highest particle
energies were expected for perpendicular shock configuration in the 
so-called "Jokipii diffusion limit", involving weakly turbulent conditions
in the large-scale magnetic field imbedded in the accreting plasma. For
such configuration we discuss the process limiting the highest energy that
particles can obtain in the first-order Fermi acceleration process due
to finite shock extensions to the sides, along and across the mean
magnetic field. Cosmic ray outflow along the shock structure can
substantially lower the upper energy limit for conditions considered for
supergalactic shocks.
     }

\maketitle
 
\section{Introduction} 

Formation of large-scale extragalactic structures (galaxy clusters and
super-clusters) involving accretion of plasma to the gravitational
potential wells is accompanied by formation of strong shock waves (Ryu
et al. 1993). These shocks were considered as viable sources of ultra
high energy cosmic rays (UHECRs), up to energies $E_{max} \sim 10^{18 -
19}$ eV (Kang et al. 1997, Hillas 1984). Evaluation of this maximum
energy is not reliable due to inadequate knowledge of physical
conditions at the shocks considered. The first estimates show that
maximal particle energies accelerated in such conditions must be limited
because of the extensive time scales involved. The magnetic fields in the
considered flows are not expected to be stronger than $\sim 1$~$\mu$G
for the accretion in rich galaxy clusters and $\sim 0.1$~$\mu$G for
supergalaxy scale accretion (Kronberg 2001). At the same time the former
flows can involve plasma velocities $\sim 1000$~km/s, while the latter a
few times lower ones (Juszkiewicz et al. 2000). The ordinary 
diffusive Fermi acceleration considered (Drury 1983) is fastest for highly
turbulent conditions with the diffusion coefficient comparable to the
Bohm diffusion coefficient, $\kappa_B$. In the considered situations, the
involved acceleration time scale $\tau \sim \kappa_B / V_{shock}^2$
becomes longer than the age of the universe for particle energies
greater than $\sim 10^{18}$ eV.

However, as pointed out by Kang et al. (1997; cf. Jokipii 1987 and
Ostrowski 1988), even faster acceleration can take place at
perpendicular shocks occurring when large-scale magnetic field
structures are present in the intergalactic medium and the accretion
proceeds perpendicular to the field direction. Then, the cosmic ray
cross-field diffusion coefficient governs the first order Fermi
acceleration process at the shock. With the involved spatial scales
{\it along the shock normal} of respectively 1 Mpc for a cluster, or a
few Mpc for a super-cluster flow, there is enough space and time to
accelerate particles up to $\sim 10^{21}$ eV, three orders of
magnitude greater than in the Bohm diffusion limit (cf.
Siemieniec-Ozi\c{e}b{\l}o \& Ostrowski 2000).

Below, in section 2, we present evaluations of the acceleration time
scales in perpendicular shocks. The derived time scale is used in
section 3 to discuss the particle diffusion and drifts along the shock
front, perpendicular to the plasma flow direction. We show that
particle escape to the sides of the shock structure may substantially
limit $E_{max}$ derived for perpendicular shocks. In section 4 we
briefly discuss the importance of this constraint for large-scale
extragalactic shocks.

\begin{figure}
\vspace{87mm}
\includegraphics{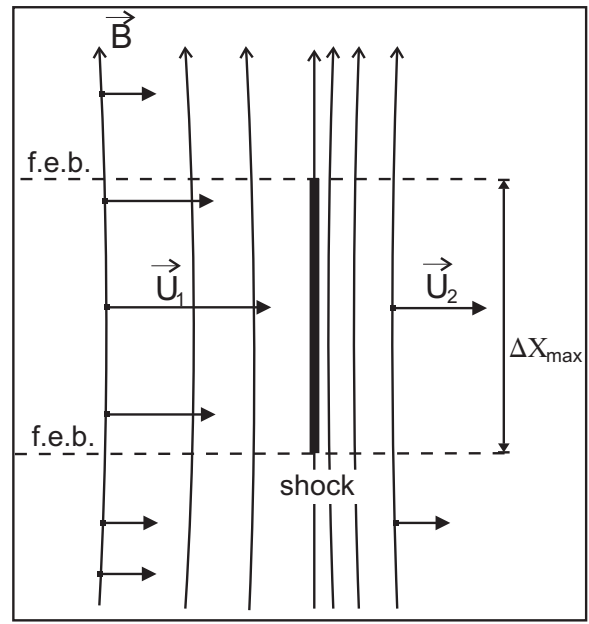}
\caption{Qualitative sketch of the physical situation considered near
the extragalactic-scale accretion shock: lines represent the mean
magnetic field which is (quasi) perpendicular to the plasma flow at the
shock. A non-uniformity of the flow results in a finite shock extension
along the mean magnetic field $\Delta X_{max}$. Particle escape along
the magnetic field  from the acceleration region at the shock is
represented by the introduced free escape boundaries (`feb') indicated
by dotted lines).}
\end{figure}

\section{Particle acceleration at perpendicular shock waves}
 
The acceleration time scale derived within the diffusion approximation
for oblique shocks (Ostrowski 1988) becomes formally invalid for
perpendicular shocks which are always superluminal. Nevertheless, Jokipii
(1987) used the diffusion equation to derive the energetic particle
acceleration time scale for such shocks:

$$\tau = {3 \over U_1 - U_2} \left( {\kappa_1 \over U_1} + {\kappa_2
\over U_2}\right) \qquad , \eqno(2.1)$$

\noindent
where $\kappa_1$ and $U_1$ ($\kappa_2$ and $U_2$) are the normal diffusion
coefficient and shock velocity as measured upstream (downstream) of
the shock. With the scaling assumed by Jokipii (1987)

$${\kappa_\perp \over \kappa_\|} = {1 \over 1 + \left( {\lambda_\| \over
r_g} \right)^2 } \qquad , \eqno(2.2)$$

\noindent
where symbols $\perp$ and $\|$ indicate quantities perpendicular and
parallel to the mean magnetic field, respectively, and $\lambda_\|$ is a
mean free path for particle scattering along the magnetic field
($\kappa_\| = {1 \over 3} \lambda_\| w$, where $w$ is a particle
velocity), and $r_g = pc/eB$ is the particle gyroradius. The equation
(2.2) describes the cross field diffusion, where, at the mean
free time, the particle moves on average $\lambda_\|$ along the magnetic
field and is scattered at a distance $r_g$ across it. Continuing the
acceleration process at perpendicular shocks requires sufficiently
effective particle cross-field diffusion. With $\eta \equiv \lambda_\| /
r_g$ this requirement can be expressed (Jokipii 1987) as

$$\eta < {w \over U_1} \qquad . \eqno(2.3)$$

\section{Maximum energy of accelerated particles}

\subsection{$\tau$: a diffusive approach}

In order to evaluate the maximum energy attainable in the acceleration
process acting at the perpendicular shock wave {\it of a finite
extension $\Delta X_{max}$ along the magnetic field} one has to consider
particle escape from the shock region, represented by free escape
boundaries of figure 1. Let us assume Eq.~(2.1) to be a valid
acceleration time estimate. Additionally, we consider conditions with
the constant $\eta $ upstream and downstream of the shock i.e. $\eta =
\lambda_{\|,1} / r_{g,1} = \lambda_{\|,2} / r_{g,2}$~. With the shock
compression ratio $r \equiv U_1 / U_2$ and a weakly perturbed upstream
magnetic field $B_1$, the downstream magnetic field $B_2 = r B_1$ and
the particle gyroradius becomes $r_{g,2} = r_{g,1} / r$~. Fast
acceleration at perpendicular shocks appears only for weakly turbulent
conditions with $\lambda_\| \gg r_g$. From Eq. (2.2), in this limit
${\kappa_\perp = \kappa_\| \left( {r_g \over \lambda_\|} \right)^2 }$.
For simplicity let us consider a purely perpendicular shock with the
mean magnetic field parallel to the shock surface and, respectively,
$\kappa_1 = \kappa_{\perp,1}$ and $\kappa_2 = \kappa_{\perp,2}$, and
relativistic cosmic ray particles with $w = c$. Then, with the Bohm
diffusion coefficient $\kappa_B \equiv {1 \over 3} r_g c$, the diffusion
coefficients

$$ \kappa_i = {1 \over \eta} \kappa_{B,i}  \quad (i\, =\, 1,\, 2) \qquad
, \eqno(3.1)$$

\noindent
and the equation (2.1) transforms to

$$\tau = {2r \over r-1} {1 \over \eta} {r_{g,1} \over c} \left( {c \over
U_1} \right)^2 \qquad . \eqno(3.2)$$

\noindent
With the constraint (2.3) and the strong shock with $r = 4$ the
acceleration time scale must be longer than $c/U_1$ times a particle
gyroperiod.

\subsection{$\tau$~: a lower limit for the perpendicular shock}

An independent lower limit for the acceleration time scale can be
derived by considering physical conditions approaching a weak scattering
limit. Then, in the absence of scattering, let us consider the
shock-drift acceleration process of upstream particles at the
perpendicular shock propagating in a uniform magnetic field. A particle
with the perpendicular momentum component $p_\perp$ compressed in the
shock will increase it to

$$p_{\perp,2} = p_{\perp,1} \, \sqrt{ r } \eqno(3.3)$$

\noindent
due to the approximate magnetic momentum conservation. For strong
shocks, on average particles roughly double their momenta at individual
transmissions through the shock, which take approximately

$$\Delta t = { 2 r_{g,1} \over U_1} = {2 r_{g,2} \over U_2} \qquad .
\eqno(3.4)$$

\noindent
Thus the characteristic acceleration time cannot be shorter than $\tau
\sim \Delta t$. The same result can be obtained by considering particle
drift along the shock, perpendicular to the weakly perturbed magnetic
field. Addition of the diffusive process that is required to allow downstream
particles to return to the shock to continue acceleration can only make
$\tau$ longer. One may note that the minimum acceleration time in the
Jokipii (1987) derivation approximately coincides with the above one.
Thus, the time (3.4) can be considered to be an absolute lower limit for
the acceleration time in the perpendicular shock and thus it defines the
ultimate limit for $\kappa_\perp$ reduction (or the upper limit for the
considered $\eta$) in derivation of the expression (3.2).

\subsection{New constraints for $E_{max}$}

Cosmic ray removal from the acceleration region near the shock proceeds
due to particle advection with the plasma flow downstream of the shock
and due to diffusion or drifts towards the escape boundaries around the shock
region. In the present paper we discuss the role of  such escape in
limiting the maximum energy of accelerated particles.

\subsubsection{A limit due to escape along the magnetic field}

First let us discuss the diffusive escape along the magnetic field
(cf. Fig.~1) limiting the maximum particle energy at the shock of a
finite extension. To evaluate this limit we consider a solution of the
1-D spatial diffusion equation for particles moving along the magnetic
field (along the $x$-axis of our coordinate frame), with the diffusion
coefficient $\kappa_\|$. For this evaluation we neglect particle
transport perpendicular to this field, both along the shock normal and
to the sides. For $N_o$ particles injected at $x = 0$ and $t = 0$, their
distribution along the $x$-axis is given by

$$n(x,t) = {N_o \over \sqrt{4 \pi \kappa_\| \, t} } \exp{\left[ -{x^2
\over 4 \kappa_\| \, t} \right] } \qquad . \eqno(3.5)$$

\noindent
If a dispersion of this distribution $\sqrt{2 \kappa_\| \, t}$ becomes
larger than the shock sideway extension $\Delta X_{max}$, the
accelerated particles tend to efficiently escape to the sides, leading
to the spectrum steepening and, finally, to formation of the energy
cut-off. Such modification of the particle spectrum is possible when the
time required for escape  becomes comparable to the acceleration time
scale $\tau$, i.e. when the particle dispersion along the $x$-axis over time $\tau$ becomes comparable to the shock extension $\Delta X_{max}$:

$$ \Delta X_{max} \approx 2 \, \sqrt{2 \kappa_\| \tau} \qquad . \eqno(3.6)$$

\noindent
Applying the above Eq. (3.2) and the expression for $\kappa_\|$, we
obtain the requirement for diffusively accelerated particles at such
finite-extent shocks as

$$r_{g} \le {\Delta X_{max} \over 2} \, {U \over c} \qquad . \eqno(3.7)$$

\noindent
As the particle gyroradius $r_{g} = pc/eB$, this expression provides the
upper energy limit for accelerated particles due to particle escape
along the mean magnetic field. For nonrelativistic shocks with the
factor $U/c \ll 1$, Eq.~(3.7) provides a severe constraint for the
maximum accelerated particle energy.

\subsubsection{A limit due to drift across the magnetic field}

As discussed in section 3.2, particle acceleration at the perpendicular
shock with the weakly perturbed magnetic field involves particle drift
along the shock surface, across the magnetic field. Thus, the shock
extension along this direction (along the $y$-axis of our coordinate frame)
also provides a limit for the maximum energy of accelerated particles.
The energy gain of a particle drifting along the shock is

$$ \Delta E = | q {\cal{E}} L | \qquad , \eqno(3.8)$$

\noindent
where the "V$\times$B" electric field $\vec{\cal{E}} = - (\vec{U}_1
\times \vec{B}_1)/c = - (\vec{U}_2 \times \vec{B}_2)/c$, $q$ is the
particle electric charge and $L$ is the distance the particle drifted
along the shock ($\Delta E > 0$). With the above expressions for the
perpendicular shock the equation (3.8) can be rewritten as

$$ {\Delta E \over E} \approx {U \over c} {L \over r_g} \qquad .
\eqno(3.9)$$

\noindent
In order to double the original energy in this process, $\Delta E = E$,
an average particle has to move a distance $L = (c/U) \cdot r_g$ along
the shock surface. Thus, if the shock extension available for drifting
particles $L = \Delta Y_{max} / 2$, the maximum energy of particles
accelerated by the shock drift mechanism is provided by the condition

$$ r_g \le {\Delta Y_{max} \over 2} {U \over c} \qquad , \eqno(3.10)$$

\noindent
analogous to the above condition (3.7).

\section{Conclusions} 
 
The maximum energy a particle can gain in the drift - diffusion
acceleration expressed by the famous Hillas (1984) formula needs more
specification concerning the characteristic spatial scales of the shock.
Apart from the shock normal extension (Bhattacharjee \& Sigl 2000) one
also needs information about its sideway extent ($\Delta X_{max}$ and
$\Delta Y_{max}$) influencing the maximum energy and, at highest
energies, the spectral shape of the accelerated particles. As an
illustration let us consider an accretion flow with velocity $U =
300$~km/s, the size of $h \approx 3$~Mpc along the flow and of $L
\approx 10$~Mpc to the sides, harbouring a shock wave inside with, say,
$\Delta X_{max}$ and/or $\Delta Y_{max} \approx L$~. In the
perpendicular shock configuration the plasma normal extent provides the
upper limit for particle energy expressed as $r_g < h$. On the other
hand particle escape to the sides provide the limits (3.7) and (3.10).
If the magnetic field in the plasma equals $0.1$~$\mu$G, the former
constraint requires approximately that $E_{max} < 10^{20}$ eV, while the
others provide the limit $E_{max} \sim 10^{17}$ eV. When varying the above accretion flow parameters within the reasonable ranges
one finds that the upper energy limit for particles accelerated at such
shocks can hardly exceed a scale of EeV, and in most cases is close to
$10^{17}$ eV. This low $E_{max}$ results from the fact that for the
weakly perturbed magnetic field required for fast acceleration at the
considered perpendicular shock, the accelerated particles can diffuse
and drift far along the finite shock to escape to its sides. For highly
turbulent conditions and/or the parallel shock configuration a similar
value for $E_{max}$ arises due to the slow acceleration constrained by
radiation losses (cf. Kang et al. 1997).

One should note that obtaining the same expressions for constraints
(3.7) and (3.10) is not a coincidence. The upper limit (3.7) can be
rewritten as $\lambda_\| \le \Delta X_{max} / 2$, while the limit (3.10)
means $L \le \Delta Y_{max} /2$~. So, for the highest energy particles
their dispersion along the shock can be at most comparable to their mean
free path or their maximum drift length. Such particles stream along the
magnetic field nearly free, advancing their positions a distance $\sim
r_g$ in each gyroperiod, and they drift across the magnetic field
changing their $y$ coordinate $\sim r_g$ at each gyroperiod, as well,
provided $B_2 \gg B_1$. Thus for a given minimum acceleration time (3.2)
or (3.4) such highest energy particles can move comparable distances
along and across the magnetic field at the shock.
 
\begin{acknowledgements}    

We are grateful to the anonymous referee for useful suggestions.
The work was supported by the {\it Komitet Bada\'n Naukowych} through
the grant PB 258/P03/99/17 (MO) and within the project 2~P03B~112~17 (GSO).
 
\end{acknowledgements} 
 
{}
 
\end{document}